\documentclass[journal,comsoc]{IEEEtran}
\usepackage[utf8]{inputenc} 
\usepackage[T1]{fontenc}
\usepackage{url}
\usepackage{ifthen}
\usepackage[cmex10]{amsmath}
\usepackage{array} 

\usepackage{enumerate}
\usepackage{amssymb}
\usepackage{amsmath} 

\usepackage{algorithm}
\usepackage{algorithmicx}
\usepackage{algpseudocode} 

\usepackage{hhline}
\usepackage{mathrsfs}
\usepackage{bm}
\usepackage{tikz}
\usetikzlibrary{arrows}
\usepackage{subfigure}
\usepackage{graphicx,booktabs,multirow}
\usepackage{epstopdf}
\usepackage{multirow}
\usepackage{tabularx} 
\usepackage{booktabs}
\usepackage{cite}

\usepackage{pifont}

\definecolor{colorhkust}{RGB}{20,43,140}
\definecolor{colortsinghua}{RGB}{116,52,129}
\definecolor{color1}{RGB}{128,0,0}

\usepackage{amsthm}

\theoremstyle{definition}

\theoremstyle{remark}

\begin{document}

        \title{Large Language Models Empowered Autonomous Edge AI for Connected Intelligence} 
      \author{Yifei Shen, Jiawei Shao, Xinjie Zhang, Zehong Lin, Hao Pan, Dongsheng Li, Jun Zhang, Khaled B. Letaief
      \thanks{This work was supported by the Hong Kong Research Grants Council under the Areas of Excellence scheme grant AoE/E-601/22-R and NSFC/RGC Collaborative Research Scheme grant CRS\_HKUST603/22. Y. Shen, H. Pan, and D. Li are with Microsoft Research Asia, Shanghai, China; J. Shao, X. Zhang, Z. Lin, J. Zhang, and K.B. Letaief are with the Department of Electronic and Computer Engineering, Hong Kong University of Science and Technology, Hong Kong.  (The corresponding author is J. Zhang). }}
        \maketitle
\begin{abstract}

The evolution of wireless networks gravitates towards connected intelligence, a concept that envisions seamless interconnectivity among humans, objects, and intelligence in a hyper-connected cyber-physical world. Edge artificial intelligence (Edge AI) is a promising solution to achieve connected intelligence by delivering high-quality, low-latency, and privacy-preserving AI services at the network edge. This article presents a vision of autonomous edge AI systems that automatically organize, adapt, and optimize themselves to meet users' diverse requirements, leveraging the power of large language models (LLMs), i.e., Generative Pretrained Transformer (GPT). By exploiting the powerful abilities of GPT in language understanding, planning, and code generation, as well as incorporating classic wisdom such as task-oriented communication and edge federated learning, we present a versatile framework that efficiently coordinates edge AI models to cater to users' personal demands while automatically generating code to train new models in a privacy-preserving manner. Experimental results demonstrate the system's remarkable ability to accurately comprehend user demands, efficiently execute AI models with minimal cost, and effectively create high-performance AI models at edge servers.

\begin{IEEEkeywords}
Edge artificial intelligence, connected intelligence, large language models, task-oriented communication, federated learning.
\end{IEEEkeywords} 
\end{abstract}

\section{Introduction}
As 5G networks are being globally deployed, researchers, corporations, and governments are exploring future visions and emerging technologies that will shape the future beyond 5G. Connected intelligence is expected to be the central focus in future networks, facilitating seamless interconnections among humans, objects, and intelligence within a hyper-connected cyber-physical world \cite{letaief2019roadmap}. Edge AI offers a promising solution for achieving connected intelligence by enabling data collection, processing, transmission, and consumption at the network edge. On one hand, advancements in wireless communication \cite{letaief2019roadmap} and low-resource machine learning \cite{kallimani2023tinyml} provide every mobile user and Internet-of-Things (IoT) device with high-quality, low-latency, and privacy-preserving AI services. On the other hand, the convergence of deep learning models and wireless networks has led to AI-enabled wireless systems with semantic communication, network sensing, cooperative perception, and intelligent resource allocation \cite{qin2021semantic,li21deep,letaief2022Edge}.

Despite these advancements, the current vision of edge AI has not yet bridged the gap to fully achieve connected intelligence. Existing studies have divided edge AI systems into several components and tackled these components separately. While designing these components is crucial, seamlessly integrating them remains a complex task. 
First, to cater to specific user demands, sensors and AI models from various vendors must collaborate in a personalized manner. Second, users may need to train customized AI models to meet particular requirements while preserving their privacy. Traditionally, this process requires interdisciplinary human experts to read manuals of edge servers, sensors, and AI models. Moreover, these experts also need to write code to parse users' requests, integrate diverse sensors, and train new AI models.
To realize connected intelligence, edge AI systems should automatically organize, adapt, and optimize themselves to meet users' diverse requirements. We refer to such systems as \emph{autonomous edge AI} systems.

Recently, LLMs have significantly influenced research and application areas due to their human-like text comprehension and generation. The LLMs' extraordinary prowess in language modeling has been leveraged in recent research for applications in wireless networks. For example, LLMs' vast stored knowledge can reduce communication costs in semantic communications \cite{jiang2023large}. Additionally, GPT-2 has been used for predicting user preferences and intents in intent-driven networking \cite{chen2023netgpt}. However, simply possessing language capabilities is not enough to achieve autonomous edge AI. To organize edge AI models, LLMs should understand the AI model's specialty by reading their manuals and making plans to invoke appropriate AI models to meet users' needs. The adaptation and optimization of edge AI models require LLMs to be proficient in coding to modify the code of AI models. Fortunately, recent works in machine learning society have shown that GPT-3.5 and GPT-4 possess capabilities in planning and coding, which is not exhibited in previous machine learning models \cite{openai2023gpt,bubeck2023sparks,zhang2023mlcopilot,shen2023hugginggpt}.

\begin{figure*}[htb]
	\centering
	\includegraphics[width=0.9\textwidth]{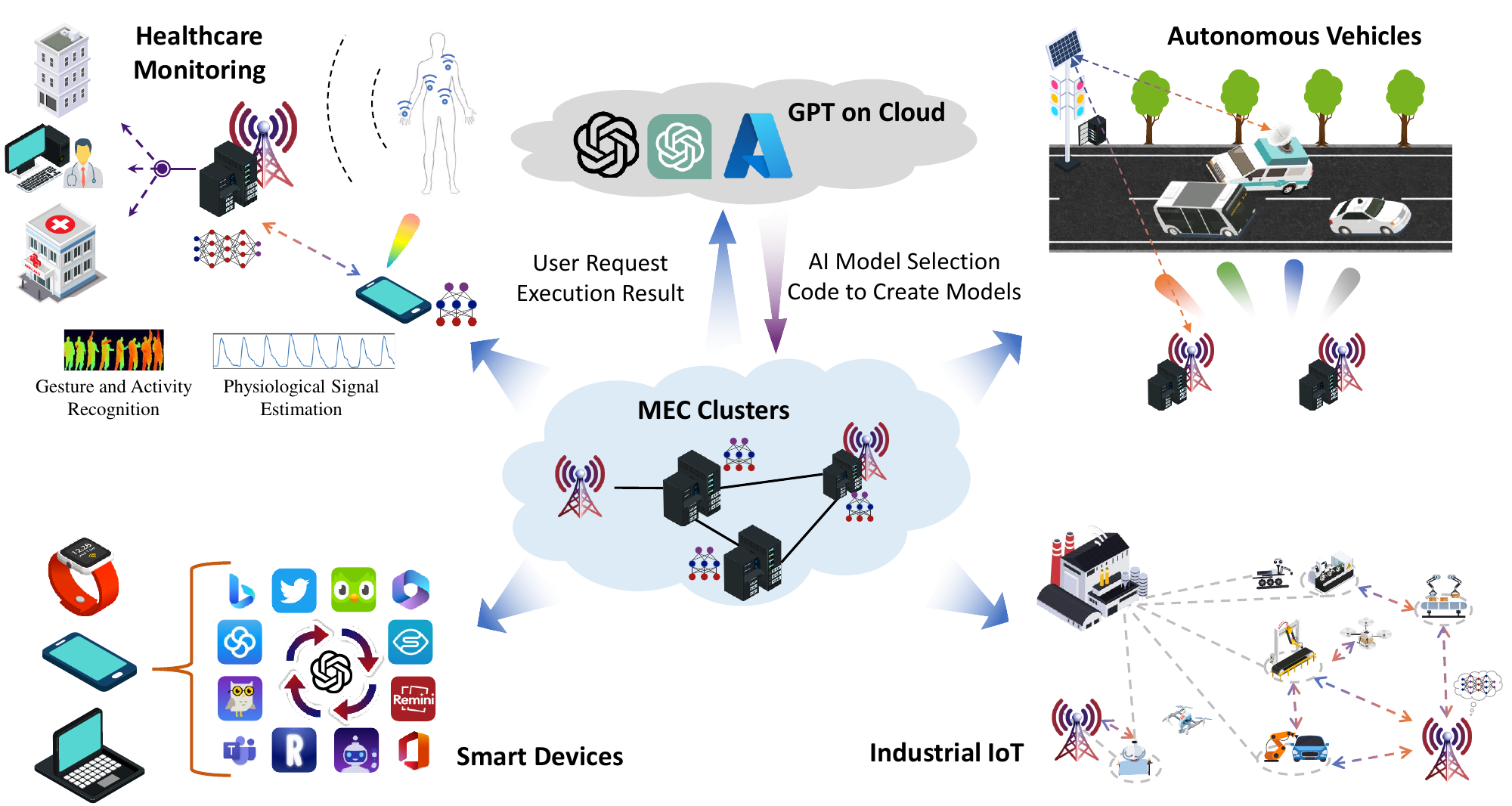}
	\caption{The vision of GPT-empowered autonomous edge AI system.}
	\label{fig:edgeai}
\end{figure*}

In this article, we explore the potential of autonomous edge AI utilizing GPT. Our framework coordinates existing edge AI models to cater to the user's personal needs and enables automatic federated learning to create new AI models with privacy considerations. Specifically, in edge AI model coordination, the requests from users are expressed in natural language, which is initially processed by GPT to understand and route them to the appropriate edge AI models. To minimize latency, we employ a task-oriented inference mechanism specifically designed for edge AI models. This approach facilitates autonomous and low-latency collaboration among edge AI models. With respect to automatic federated learning, we first prepare a template code and identify the key configurations to be modified, based on which GPT generates the detailed configurations and writes the corresponding code. This method presents an automated approach for crafting deep learning models that adhere to privacy constraints. Our contributions can be summarized as follows:

\begin{enumerate}
    \item We present a vision of autonomous edge AI and explore the potential of utilizing GPT to achieve this objective. We also discuss the unique advantages and challenges it presents.
    
    \item We propose a cloud-edge-client hierarchical framework for autonomous edge AI, which integrates the powerful ability of GPT to understand users' intentions and generate code, as well as the classic wisdom from edge AI to reduce latency and increase reliability.
    
    \item We implement a proof-of-concept system. The experimental results show the system's capability to accurately comprehend user demands, execute AI models with minimal delay, and effectively create high-performance AI models.
    
\end{enumerate}

\section{Autonomous Edge AI: Challenges and Opportunities}
In this section, we first present the concept of autonomous edge AI, and then explore the potential of GPT as an enabler for autonomous edge AI, emphasizing its distinct capabilities.

\subsection{From Edge AI to Autonomous Edge AI}
As data continues to proliferate, the demand for AI services increases correspondingly. Edge AI has emerged as a key technology due to its capacity to support ubiquitous AI services. Several methods have been developed for effective, secure edge training and inference \cite{letaief2022Edge}. For instance, federated learning \cite{mcmahan2017communication} trains a global statistical model without accessing raw data from edge devices, thereby paving the way for private and secure edge training. Semantic communication \cite{qin2021semantic} marks a paradigm shift in communication system design from data recovery to task accomplishment, facilitating communication-efficient edge inference for big data. TinyML \cite{kallimani2023tinyml} enables low-cost, low-power devices to execute AI models, thereby contributing to the deployment of AI models on IoT devices.

Thanks to these advancements, numerous sensors and AI models will be deployed at edge servers in future wireless networks. However, the users, vendors, and developers are faced with the following distinct challenges:

\begin{enumerate}
\item \textbf{User interface design:} Traditionally, users interact with AI models through apps, necessitating their effort to familiarize themselves with each app's functionality. The user experience could be significantly enhanced by employing an intelligent dispatcher that routes user requests in natural language to appropriate apps, taking into account user preferences and context.

\item \textbf{Sensor and AI model integration:} To satisfy user demands, vendors must integrate sensors and AI models from various providers, each possessing unique capabilities and strengths. Achieving seamless coordination between multiple sensors and AI models has traditionally demanded significant efforts from support teams to design standardized protocols. Furthermore, the introduction of new edge devices necessitates support teams to understand the devices' manuals and integrate them into a unified protocol.

\item \textbf{AI model customization:} The developers or users may need to train customized AI models to meet specific requirements, posing two main challenges. First, human experts with interdisciplinary expertise in machine learning and the application domain to write the code are required. Second, experts must prioritize user privacy during the training process by utilizing techniques such as federated learning to ensure data confidentiality. 

\end{enumerate}

In the current vision of edge AI \cite{letaief2022Edge}, these challenges are either unexplored or left to human experts from the operation team to address manually. However, in the realm of connected intelligence, we anticipate an edge AI system that is capable of automatically organizing and enhancing AI models to cater to users' demands and intelligently allocate resources to meet network constraints, which we refer to as \emph{autonomous edge AI systems}. This is done by creating an edge AI system endowed with self-organization and self-improvement capabilities.

The \emph{self-organizing ability} refers to efficiently orchestrating the existing sensors and AI models at the edge servers based on user requests in natural languages. Specifically, it encompasses comprehending user requests in natural languages, allocating these requests to a network of sensors, executing AI models on the sensors efficiently, and consolidating the AI model outcomes to accomplish the task, ultimately addressing sensor integration and user interface design challenges. Existing edge AI frameworks primarily concentrate on achieving efficient inference of AI models \cite{letaief2022Edge}, leaving other components largely unexplored. The \emph{self-improvement ability} involves generating code to fine-tune existing AI models and training new ones in a privacy-preserving manner, thereby tackling AI model customization challenges. While current studies primarily focus on algorithm design for communication-efficient and private edge learning \cite{letaief2022Edge}, the process of code implementation still necessitates the involvement of human experts.

Autonomous edge AI unveils a plethora of possibilities across diverse applications. In healthcare monitoring, such a system empowers real-time tracking and analysis of patients' vital signs, pinpointing potential health hazards and proposing tailored preventative actions based on individual needs and conditions. In autonomous driving, it streamlines the integration of AI models in charge of perception, decision-making, and control, thereby guaranteeing safety, efficiency, and adaptability. In industrial IoT, the system autonomously generates code to modify existing AI models and develop new ones tailored to the distinct requirements of various industrial settings. In smart devices, the system delivers context-aware recommendations, automates routine tasks expressed in natural language, and improves the overall user experience. The vision of autonomous edge AI is depicted in Fig. \ref{fig:edgeai}.

To sum up, the autonomous edge AI system possesses immense potential to transform a wide range of wireless applications. Nonetheless, it demands a powerful controller to handle users' diverse and personal requests, clear understanding of each sensor and AI model's unique characteristics, along with the capability to produce code. In the following subsection, we delve into the prospect of utilizing GPT as the controller.

\subsection{GPT as a Potential Solution}\label{sec:GPT}
OpenAI recently released a powerful large language model named GPT \cite{openai2023gpt}. Thanks to the extensive training data in language, mathematical reasoning, and programming code, GPT demonstrates an impressive ability to handle complex tasks in fields such as mathematics, coding, medicine, law, and psychology \cite{bubeck2023sparks}. GPT possesses several unique characteristics not found in previous machine learning models. These abilities play a vital role in autonomous edge AI systems, as detailed below.

\begin{enumerate}
\item \textbf{Language understanding:} Among GPT's most significant competencies is its ability to answer open-domain questions spanning various subjects. In autonomous edge AI, this ability enables GPT to understand users' intentions and recognize the specialties of edge AI models by comprehending their manuals.

\item \textbf{Planning ability:} A key feature of GPT is its aptitude for planning. It can assess a given situation, pinpoint potential sub-tasks, and formulate a plan to achieve a desired outcome. In autonomous edge AI, this ability helps GPT plan the solvable sub-tasks based on the users' ambiguous requests and dispatch the sub-tasks to edge AI models.

\item \textbf{In-context learning:} GPT is the first model capable of learning from provided knowledge and examples within a context (or prompt), without requiring any fine-tuning. In autonomous edge AI, information about devices and edge AI models might not be included in GPT's training dataset, but can be supplied in this manner.

\item \textbf{Code generation:} A notable advantage of GPT is its ability to understand and generate code. In autonomous edge AI, we utilize this capability to refine existing AI models and create new ones. 
\end{enumerate}

In summary, GPT can understand users' intentions based on its language understanding ability and dispatch user requests to edge AI models using its planning ability. This lays the groundwork for coordinating sensors and AI models at the edge server, exemplifying the self-organizing ability. Additionally, GPT's capacity to generate code for training new AI models sets the stage for its self-improvement ability. Therefore, GPT holds significant potential for achieving autonomous edge AI. In the next section, we will elaborate on our proposed framework.

\begin{figure*}[htb]
	\centering
	\includegraphics[width=1\textwidth]{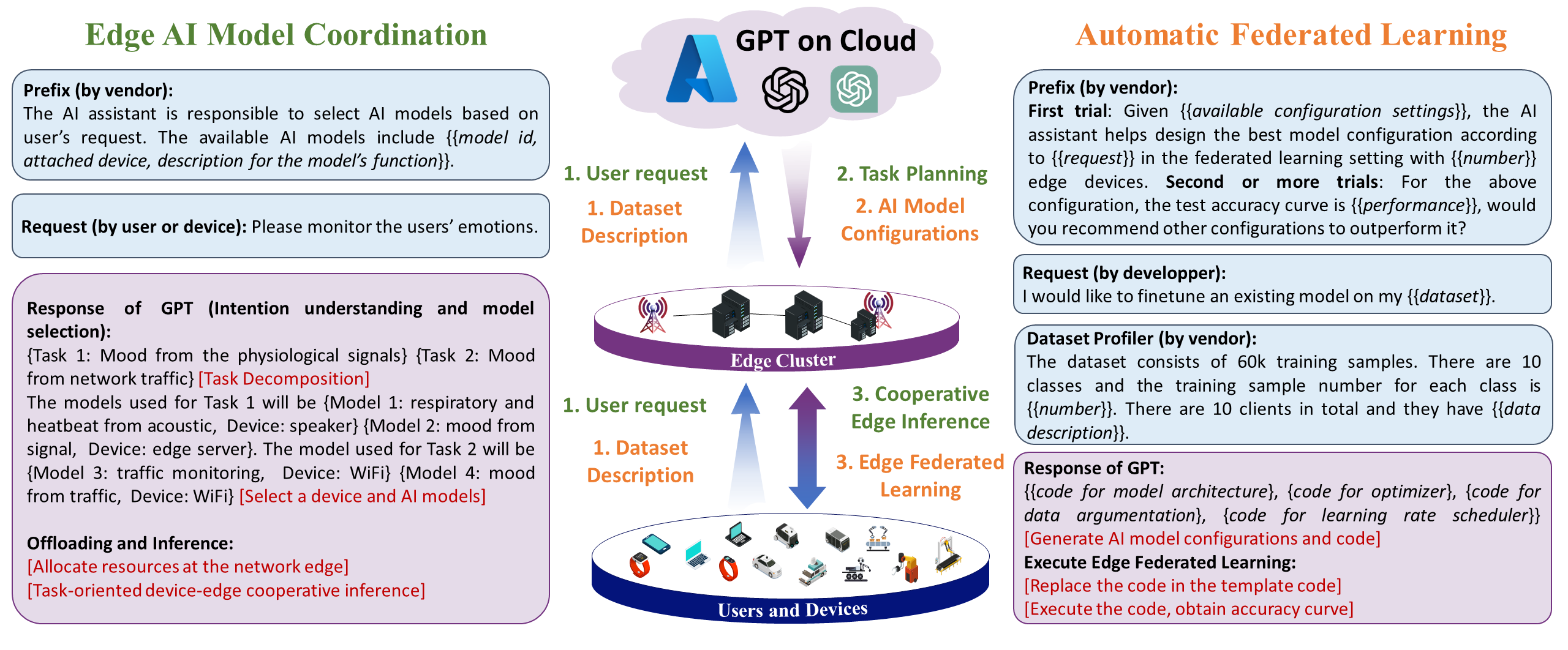}
	\caption{The overall framework of autonomous edge AI: automatically coordinating AI models at the edge and generating federated learning code to update or train new AI models.}
	\label{fig:edgegpt}
\end{figure*}

\section{Cloud-Edge-Client Hierarchical Framework for Autonomous Edge AI}
In this section, we introduce a framework for achieving autonomous edge AI in the cloud-edge-client hierarchical system. Cloud servers, e.g., Microsoft Azure, boast substantial computational resources to support GPT. The client layer consists of users' devices and IoT sensors, which possess limited computational resources. Edge servers, strategically located near the clients, provide enhanced computational resources. The following subsections present a detailed framework for edge AI model coordination that aims at enabling self-organization ability and automatic federated learning to attain self-improvement ability under privacy constraints.

\subsection{Edge AI Model Coordination}
This subsection presents our framework for coordinating edge AI models to address users' requests in natural language. It includes using GPT to dispatch users' requests to edge AI models and adopting task-oriented inference to reduce inference latency.

\subsubsection{User Intention Understanding and Model Selection}\label{sec:model_select}
We begin by describing how to coordinate AI models so that they can accommodate users' diverse demands that are expressed in natural language. The first step entails understanding users' intention and decomposing the requests into solvable tasks by specialized AI models. The context provided to GPT consists of the \emph{prefix} and \emph{request}. The prefix, including a pool of available AI models, is supplied by the vendor. The request, describing the users' intention in natural language, is provided by the users. With the provided context, GPT processes the prefix, request, and historical requests, ultimately selecting a series of AI models to be executed.

We then present a demonstration example of indoor wireless sensing for healthcare monitoring. The pool of available AI models is presented in the following format: \{AI model name; associated equipment; description of its function\}. An example would be \{heartbeat model; speaker; monitoring the heartbeat using acoustic signals\} or \{signal-to-mood; edge server; classifying mood based on physiological signals\}. GPT can glean information about the system from these descriptions by leveraging its in-context learning ability. Assume that the user's request is to monitor emotions. Upon receiving the request, GPT breaks it down into two sub-tasks: mood classification based on network traffic and mood detection based on users' physiological signals. Then, the first sub-task is dispatched to the traffic analysis AI models on WiFi, which detects app usage from the network traffic and employs a pre-trained decision tree to classify mood based on the app usage. The second sub-task is dispatched to the sensing AI models on the speaker, which monitors the heartbeat and respiration from acoustic signals and then applies a pre-trained random forest to detect mood based on these signals. After obtaining these two classification predictions, GPT combines them to get the final results. The process is illustrated in Fig. \ref{fig:edgegpt}.

\subsubsection{AI Model Offloading}
Upon selecting the AI models, the next step is to execute them efficiently. Traditional methods deploy AI models at a central server, which typically has access to more abundant energy supplies and computational resources than low-end devices. Nevertheless, if radio resources at the network edge are limited or the channel experiences dynamic fluctuations, sending such a large volume of raw data to the server results in excessive communication overhead.

To achieve low-latency inference, we adopt the mobile edge computing framework \cite{mao17mec}, which offloads a portion of the computation workload from sensors to nearby edge servers, enabling client-edge co-inference. By leveraging the layer-wise structure of deep neural networks, we can choose a partition point to divide the model into two parts: a lightweight model deployed on the device, and another part, typically larger and requiring more computation, deployed on the edge server. The intermediate features generated by the on-device model are sent to the edge server for further processing.

\subsubsection{Task-Oriented Compression}

To further reduce transmission latency during co-inference, we apply compression techniques to intermediate features before transmission. Numerous handcrafted and DNN-based coding algorithms in the literature lossily compress data to reduce size while maintaining acceptable distortion after reconstruction. 
However, these coding schemes are less effective because they also recover noise and redundant information in features, which hold no value for the server-side model on the receiving end. The task-oriented semantic communications have been recently proposed to resolve this issue \cite{shao20communication,chaccour2022less,qin2021semantic}.

We employ a task-oriented compression method \cite{shao20communication} for co-inference. Specifically, an entropy model is attached to the partition point for feature compression.
This model filters out the redundancy and encodes only task-relevant information into bitstreams.
The entropy model is trained using backpropagation, with the weighted sum of the inference loss and communication cost as the objective function. 
In this manner, the task-oriented communication scheme can reduce the communication overhead while preserving inference performance.

\subsection{Automatic Edge Federated Learning}
The procedures mentioned above are applicable when the necessary models are readily available on edge servers. However, there are instances where users need to customize AI models on their datasets to cater to specific needs. These datasets may contain private information, and sharing them is not advisable. Federated learning provides a solution by keeping the training data on individual devices and communicating a shared model across distributed devices, thereby enhancing data privacy and security. Traditionally, designing and executing federated learning entails human experts who have to manually design, program, and debug federated learning models. To enable autonomous edge AI, we employ GPT to automate this process, reducing the need for human intervention.

However, creating code from the ground up is a very challenging task, and the current version of GPT struggles with it. To mitigate this issue, we first prepare a federated learning template code and specify key configurations. In this article, model architectures, optimizers (including learning rate, global and local learning rate decay), and data augmentation methods are defined as key configurations. We also specify the range of template code that each configuration corresponds to. 

As elaborated in Section \ref{sec:model_select}, the context delivered to GPT includes the prefix and the request. The prefix includes the name of the configurations and the profiles of the dataset (e.g., the number of classes). The user request includes the purpose of the model (e.g., for image classification). Then GPT's response will include its suggestions for these configurations. Then the part of template code corresponding to these configurations is provided to GPT for modification. After getting GPT's response code, we have a shell script to modify the template code. Then the modified code is executed at the device and edge servers to obtain the learning curve.

To enhance the performance, we repeat this process for a few times (e.g., $3$ times). In the second or more trials, GPT then utilizes the learning curve to determine the key configurations, excluding the architecture. We empirically find that the configurations generated by GPT can improve with more iterations. The whole process is illustrated in Fig. \ref{fig:edgegpt}.

\begin{table}[t]
\caption{Evaluation for task planning. The best results are in bold. Acc and F1 represent the accuracy and F1-score, respectively.} 
\centering
\begin{tabular}{c|ccc}
\hline
Model            & Acc$\uparrow$ & F1$\uparrow$ & Latency$\downarrow$ \\ \hline
GPT-3 350M & 32.93\%  & 33.95\%  & 0.37 sec           \\
GPT-3 6.7B & 40.24\%  & 42.31\%  & 0.55 sec           \\
GPT-3 175B & 68.89\%  & 74.70\%  & 0.45 sec           \\
GPT-3 175B IT & \textbf{84.44\%}   & \textbf{85.39\%}  & 0.58 sec           \\ \hline
Zero-shot Classification & 36.59\%  & 36.59\%  & \textbf{0.28 sec}  \\ \hline
\end{tabular}
\label{Table:task_planning_results}
\end{table}

\begin{table*}[t]
\centering
\caption{The communication cost and performance of different methods in edge inference. The best results are shown in bold.} 
\resizebox{0.85\textwidth}{!}{

\begin{tabular}{c|cc|cccc|ccc}
\hline
\multirow{2}{*}{Method}                                                                         & \multicolumn{2}{c|}{Image classification} & \multicolumn{4}{c|}{Image caption} & \multicolumn{3}{c}{Visual question answering}   \\
                                                                                                &  Cost$\downarrow$             & Accuracy$\uparrow$           & Cost$\downarrow$  & BLEU$\uparrow$   & CIDEr$\uparrow$  & SPICE$\uparrow$ & Cost$\downarrow$ & Test-dev$\uparrow$ & Test-std$\uparrow$ \\ \hline
\begin{tabular}[c]{@{}c@{}}Edge-only inference with\\lossless data compression\end{tabular} & 224.41 KB                    & \textbf{84.16\%}               &  249.40 KB       & \textbf{39.66}  & 133.25 & \textbf{23.77} & 372.58 KB        & \textbf{78.24}             & \textbf{78.32}             \\ \hline
\begin{tabular}[c]{@{}c@{}}Edge-only inference with\\lossy data compression\end{tabular}    &  33.86 KB                   & 82.83\%               &   19.16 KB       & 38.84  & 129.30 & 23.32 &  20.51 KB       & 77.22             & 77.38             \\ \hline
Client-edge co-inference  &  \textbf{32.83 KB}                   & 84.02\%               &  \textbf{18.77 KB}        & 39.58  & \textbf{133.29} & 23.75 &  \textbf{20.39 KB}       & 78.22             & 78.22             \\ \hline
\end{tabular}
}
\label{Table:inference_results}
\end{table*}

\section{Experiments of Autonomous Edge AI}
This section provides performance evaluations to demonstrate the capabilities of autonomous edge AI. 

\subsection{Edge AI Model Coordination}

We first test the AI model coordination ability of the proposed framework. The evaluation includes the accuracy of model selection and the latency of the proposed system.

\textbf{Experimental setting:} We consider a cloud-edge-client hierarchical system. Specifically, the client has a single device that is a Jetson Nano board, the edge server is equipped with a GeForce RTX 4090, and the cloud server is hosted on Microsoft Azure. This system is responsible for processing users' daily photos based on their commands described by natural language. The available tasks include image classification, image captioning, and visual question answering (VQA), with each task corresponding to one expert model. Specifically, we adopt the Vision Transformer (ViT) for image classification and utilize bootstrapping language image pre-training (BLIP) to address image captioning and VQA tasks. GPT is hosted on the cloud server while other AI models are deployed at the edge and client.

\textbf{Evaluation Results:} We have curated a challenging user request dataset consisting of $82$ user requests, with each task containing $26$, $25$, and $31$ distinct requests, respectively. 
The test performance of user intention understanding and AI model selection is shown in Table \ref{Table:task_planning_results}. 
We use GPT-3 with an increasing number of parameters. 
Specifically, \emph{GPT-3 350M}, \emph{GPT-3 6.7B}, and \emph{GPT-3 175B} represent GPT-3 with 350 million, 6.7 billion, and 175 billion parameters, respectively. \emph{GPT-3 175B IT} is a variant of GPT-3 with instruction fine-tuning to align with human values. We do not consider GPT-4 as its latency is high (5s-20s). 
We also include a baseline to compare the GPT-based approach with state-of-the-art zero-shot classification provided by OpenAI.
We observe that smaller models and zero-shot classification fail to achieve satisfactory accuracy, which indicates the necessity of using GPT with a large number of parameters.

Upon selecting a specific model, the next step is to execute the task.
We compare our proposed client-edge co-inference scheme with two baselines: edge-only inference and client-only inference.
In edge-only inference, the expert models are deployed at the edge server, and the user input data are transmitted to edge servers for remote processing.
The data undergo either lossless or lossy compression methods before transmission.
Table \ref{Table:inference_results} presents the communication cost and performance of various inference schemes.
Our client-edge co-inference scheme achieves comparable performance to the edge-only inference with lossless data compression, while significantly reducing the communication cost. 
Lossy compression can also reduce the communication overhead of the edge-only inference scheme, but it results in performance degradation.
In client-only inference, the expert models are deployed at the client to locally execute the task.
While this scheme avoids high transmission delay, it may result in increased computation latency due to the limited resources available on the device.

\begin{figure}[t]
    \centering
    \includegraphics[width = 0.9\columnwidth]{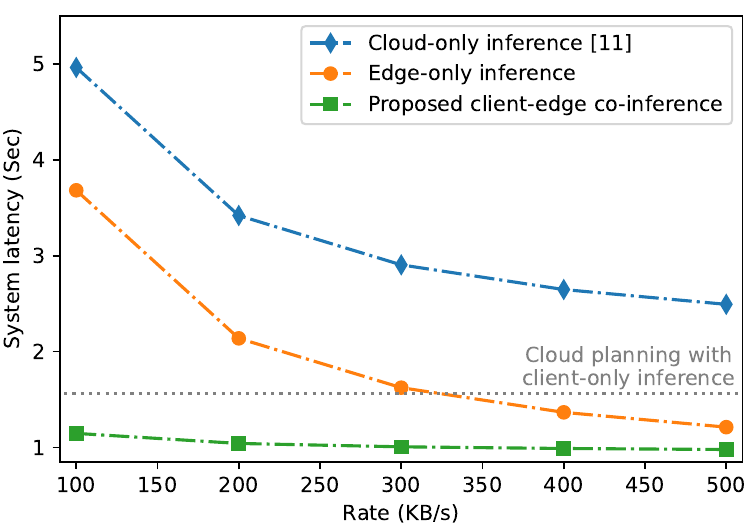}
    \caption{System latency as a function of communication rate.}
    \label{fig:system_latency}
\end{figure}

Next, we compare the end-to-end system latency of various inference schemes. The end-to-end latency includes transmission time, computation time, and downloading time. We assume that the proportion of user requests for the classification, image captioning, and VQA tasks are 10\%, 40\%, and 50\%, respectively.
The test images are from the ImageNet validation set, the COCO Karpathy split, and the VQA v2.0 dataset.
To maintain the desired performance in edge-only inference, the client's images are compressed in a lossless manner. Besides, we present the latency of the cloud-only inference method, where the cloud server performs both task planning and execution, which is the paradigm for GPT-plugins or HuggingGPT \cite{shen2023hugginggpt}. Fig. \ref{fig:system_latency} depicts the system latency under different communication rates.
It shows that cloud-only inference experiences much higher latency than other methods.
This is mainly due to the long distance between the cloud server and the client, resulting in an increased round-trip time for data transmission.
Besides, edge-only inference demonstrates comparable or even lower latency than client-only inference at high communication rates, typically ranging from 300 KB/s to 500 KB/s.
However, as the rate decreases, the latency of edge-only inference increases dramatically.
In contrast, the latency of the client-edge co-inference scheme remains consistently lower than both the edge-only and client-only inference schemes.

To sum up, these experiment results demonstrate that the proposed framework is able to understand the users' requests and execute the edge AI models to cater to diverse requests with low latency.

\begin{figure}[t]
    \centering
    \includegraphics[width = 0.9\columnwidth]{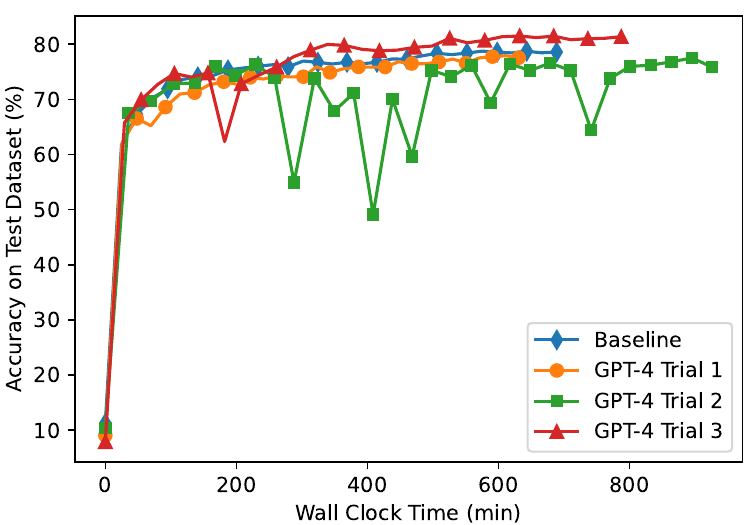}
    \caption{Results for automatic edge federated learning.}
    \label{fig:auto_fl}
\end{figure}

\subsection{Automatic Federated Learning}
This subsection evaluates the proposed framework's automatic federated learning ability among edge devices.

\textbf{Experimental settings:} We employ a classic federated averaging algorithm \cite{mcmahan2017communication}, and use GPT-4 as the scheduler, as it is highly effective in generating code \cite{openai2023gpt}. We consider 10 clients, each equipped with a device of Jetson Nano board, with independent and identically distributed (IID) partitioned CIFAR10 data. The uplink rate is 250 KB/s, and the downlink rate is 500 KB/s. The batch size at the client is 100, the local epoch is $10$, and the global epoch is 30. The users' request is to choose a model from the torch vision library, adapt it to CIFAR10 (create a new classifier), and set up the learning rate, data augmentation method, and learning rate scheduler. 

\textbf{Evaluation results:} The performance of the proposed method is shown in Fig. \ref{fig:auto_fl}. For comparison, we consider a baseline that finetunes ResNet18 with $11$ million parameters using the stochastic gradient descent (SGD) optimizer, with the learning rate determined by grid search. In the first trial, GPT suggests using SGD to fine-tune MobileNetV2, which has $3.4$ million parameters. The code for MobileNetV2 is generated by GPT, loading the pretrained model weights from the torch vision library but replacing the last layer with a randomly initialized linear function. The template code is then modified accordingly and executed. As shown in Fig. \ref{fig:auto_fl}, the performance of the first trial is inferior to the baseline, since MobileNetV2 is much smaller than ResNet18. In the second trial, the accuracies from the first trial are reported to GPT-4, which then suggests a different data augmentation method and using the Adam optimizer. However, the resulting curve oscillates as the learning rate is too large. In the third trial, GPT suggests employing a learning rate scheduler that decreases the learning rate to 0.1 of the original value every 10 global updates, which yields better performance than the baseline. In the fourth trial, it does not generate any new configurations, so we terminate the process. This experiment demonstrates that GPT is able to generate code for training federated learning models and iteratively improving the performance automatically, thus paving the way for an autonomous edge AI system that can automatically improve itself while simultaneously preserving users' privacy.

\section{Conclusions and Future Works}
This article has explored autonomous edge AI in a cloud-edge-client hierarchical system by leveraging large language models. We proposed a framework to orchestrate existing AI models and train new ones at the network edge. Sample experiments of the proof-of-concept model have validated the effectiveness of the proposed framework. Several areas could be explored for future advancements.

\textbf{Energy efficiency and privacy:} Our initial framework prioritizes latency but energy efficiency and privacy are also important in sustainable edge AI. To enhance energy efficiency, each stage in the framework can be modified to take energy consumption into consideration. In terms of privacy, we can integrate techniques such as information bottlenecks and adversarial training to filter out privacy-sensitive information.

\textbf{GPT-based scheduling algorithm selection:} In wireless communication research, numerous optimization-based and AI-based scheduling algorithms have been developed. We have demonstrated that GPT can select AI models aligned with user intentions. We expect GPT could similarly select resource allocation algorithms, tailored to the environment, to decrease latency and enhance spectral efficiency.

\textbf{Incorporating other Generative AI models:} This work centers on using LLMs within edge AI systems. However, exploring other Generative AI models is also promising due to their data compression and generation capabilities. For instance, in edge AI coordination, Generative AI could improve compression in task-oriented communication. Additionally, Generative AI has the potential to generate datasets for federated learning, offering better privacy protection.

\bibliographystyle{ieeetr}
\bibliography{ref}

\noindent{\bf{Yifei Shen}} [S'18-M'22] (yifeishen@microsoft.com) currently works at Microsoft. \\
{\bf{Jiawei Shao}} [S'20] (jiawei.shao@connect.ust.hk) is currently pursuing a Ph.D. degree at HKUST. \\
{\bf{Xinjie Zhang}} [S'21] (xinjie.zhang@connect.ust.hk) is currently pursuing a Ph.D. degree at HKUST. \\
{\bf{Zehong Lin}} [S'17-M'22] (eezhlin@ust.hk) received his Ph.D. degree from the CUHK. He is currently a Research Assistant Professor at HKUST. \\
{\bf{Hao Pan}}  (panhao@microsoft.com) is currently a Researcher at Microsoft Research Aisa. \\
{\bf{Dongsheng Li}} [M'15-SM'23] (dongsli@microsoft.com) received his Ph.D. degree from Fudan University. He is currently a Principal Research Manager at Microsoft Research Aisa. \\
{\bf{Jun Zhang}} [S'06-M'10-SM'15-F'21] (eejzhang@ust.hk) received his Ph.D. degree from the University of Texas at Austin. He is currently an Associate Professor at HKUST. \\
{\bf{Khaled B. Letaief}} [S'85-M'86-SM'97-F'03] (eekhaled@ust.hk) is a Member of the United States National Academy of Engineering, Fellow of IEEE, Fellow of Hong Kong Institution of Engineers, and Member of the Hong Kong Academy of Engineering Sciences. He received the BS, MS, and Ph.D. Degrees in Electrical Engineering from Purdue University, in 1984, 1986, and 1990, respectively. He is the New Bright Professor of Engineering at HKUST.

\end{document}